# UML4IoT - A UML profile to exploit IoT in cyber-physical manufacturing systems

Kleanthis Thramboulidis, Foivos Christoulakis
Electrical and Computer Engineering
University of Patras, Greece
*thrambo@ece.upatras.gr*

*Abstract*—Internet of Things is changing the world. The manufacturing industry has already identified that the IoT brings great opportunities to retain its leading position in economy and society. However, the adoption of this new technology changes the development process of the manufacturing system and raises many challenges. In this paper the modern manufacturing system is considered as a composition of cyber-physical, cyber and human components, and IoT is used as a glue for their integration as far as it regards their cyber interfaces. The key idea is a UML profile for the IoT with an alternative to apply the approach also at the source code level specification of the component in case that a UML design specification is not available. The proposed approach, namely UML4IoT, fully automates the generation process of the IoT-compliant layer that is required for the cyber-physical component to be integrated in the modern IoT manufacturing environment. A prototype implementation of the myLiqueur laboratory system has been developed to demonstrate the applicability and effectiveness of the UML4IoT approach.

## I. INTRODUCTION

Manufacturing systems independent of their nature address the challenge of satisfying product customization needs. Customers are expecting to have products that will address their specific needs and will be comparable in cost to mass-produced ones [1]. Discrete process control systems, such as assembly systems [48], or batch process control systems should gradually be transformed to highly adaptive and resource-efficient systems able to address the always increasing needs of product customization [2]. The industry has to address many challenges in order to successfully switch to this level of flexibility and retain its leading position in economy [3]. Multidisciplinary areas such as mechatronics and cyber-physical systems (CPS) as well as IT technologies such as Internet-of-Things (IoT) and cloud computing are playing a leading role in this industrial revolution, which is known as the fourth industrial revolution or Industry 4.0 [4]. Cyber-physical systems play an important role towards Industry 4.0. Based on a very short definition given in [5] the orchestration of the computational and physical processes that constitute the manufacturing system can be considered as a cyber-physical system. The great impact of CPSs in manufacturing based on a number of explorative case studies is examined in [6].

The traditional approach in the development of manufacturing systems considers (a) the system as a composition of the physical plant, the network of computation nodes and the computational processes required to monitor and control the physical ones, and (b) the development processes of each one of these three disciplines independent of the others with their own specific methods and tools. This approach is unable to address the demand for synergetic mechatronic dependability predictions [7] and is considered inappropriate to address the increased requirements for flexibility and evolvability of today's systems [8][9]. It does not force an actual cooperation in the development of the three discipline parts; thus it leads to a high couple between the physical parts (plant) with the corresponding parts of the cyber world (computational part).

Model Integrated Mechatronics [9] enhanced with the 3+1 SysML-view model [10] addresses this challenge by considering the system as a composition of well defined reusable mechatronic components. It proposes the tight integration of the physical world with the cyber one at the component level leading to highly cohesive components with well defined interface and behavior. This approach greatly reduces the coupling between the system components compared to the traditional one. The so created cyber-physical component, which is called mechatronic component (MTC), is composed of highly coupled mechanics, electronics and software parts to accomplish a specific need and offer higher level functionality compared to one offered by the physical unit. In this way computing and communication capabilities have been embedded in the physical components transforming these to cyber-physical components such as the ones of energy systems mentioned in [11]. This approach has already found the road to production in industry in the context of Industry 4.0, e.g., FESTO [12]. The interface of a MTC is composed of physical, cyber-physical and cyber ports through which it is integrated with other components so as to effectively collaborate with these to accomplish a higher layer of behavior that is required at the sub system or system level. The integration process of the constituent components of cyber-physical systems is a great challenge since it directly affects quality properties of the system such as adaptability, flexibility and customization.

Technologies such as the Internet of Things (IoT), Cloud computing, Service Oriented Architectures (SOA) and mobile computing if successfully adapted to the industrial automation domain may address challenges in modern manufacturing. Web standards such as SOAP and WSDL have been already adopted by research groups in the industrial automation domain and several approaches have been described to exploit their benefits, as for example [13-15] to mention a few. SOA based products have already appeared in the industrial systems market in the context of Industry 4.0. For example, TwinCAT from Beckhoff combines IEC 61131-3-based SOA services with OPC UA interoperability [16]. However, technologies such as SOAP and WSDL, have been proved too heavyweight compared to the recent IoT protocol stacks. On the other side IoT is aligned well with the architecture of a manufacturing enterprise and as authors argue in [3] it is able to provide "vital solutions to planning, scheduling, and controlling of manufacturing systems at all levels." IoT brings great opportunities in achieving better system performances in globalized and distributed environments. However, as authors claim in [3], IoT in manufacturing is in infant stage and there is a demand for research, development and standardization of enabling technologies for safe, reliable, and effective communication and decision-making. There is a need for platforms to provide information integration, repository services and support for analysis of the whole IoT-based system [17]. The effective exploitation of IoT in the domain of cyber-physical manufacturing systems is a challenge for the academy and industry.

The approach presented in this paper effectively integrates trends in cyber physical systems and IoT and describes a framework that address challenges introduced by the use of IoT in the development process of manufacturing systems. It automates the generation process of the IoT-compliant layer for new mechatronic components but also for legacy ones to exploit the IoT connectivity. Two alternatives are presented and discussed. The first one is based on the UML design specification of the cyber part of the mechatronic component; the second one is based on the source code if a higher level design specification such as the UML one is not available. Java is used as a case study but other languages, such as the IEC 61131, can also be considered.

The presented approach integrates modeling techniques required for the specification of complex cyber physical components with IoT technologies. More specifically, it exploits the OMA LWM2M application protocol [18] and IPSO smart objects [19]. LWM2M and IPSO smart objects focus on modeling the exposed interface of simple smart objects and are not able to address the modeling needs of complex components of manufacturing systems. Thus, the IPSO smart objects model is adopted to model only the exposed by the component interface and transform the component to an IoT-compliant one. SysML and UML are utilized as described in [20] for the modeling of the mechatronic component. However, extensions are proposed so as to enable an automatic generation of the IoT-compliant interface of the component that transforms it to an Industrial Automation Thing. These interfaces, if properly used at the system or subsystem integration level may lead to on demand system configurations that address specific customer needs in a cost effective way. The main contributions of this paper are: (a) the definition of a UML profile for the IoT, namely the UML4IoT profile, (b) the automation of the generation process of the IoT-like interface i.e., of the IoT wrapper, of the mechatronic component, and (c) a lightweight flexible prototype implementation of the OMA LWM2M protocol based on meta programming.

The remainder of this paper is structured as follows. In the next section the proposed in this paper approach, namely UML4IoT, as well as the example system used as case study are briefly presented. The UML profile for IoT and its exploitation to automate the generation process of the IoT wrapper of the mechatronic component is presented in Section III. The process for the automation of the generation of the IoT-compliant smart object is described in Section IV. In Section V, related work is presented. Evaluation and measurements on the prototype implementation of the example system used as case study are given in Section VI and the paper is concluded in the last Section.

## II. AN OVERVIEW OF THE UML4IoT APPROACH

### A. The myLiqueur production system

The liqueur plant system used as case study in [21] was adopted as base to define the myLiqueur production system, which exploits IoT to allow end users to produce custom types of liqueur. Production parameters that define the specific type of liqueur could be defined by the end user through the myLiqueur App. The myLiqueur production system is composed of the following mechatronic components, as shown in Fig. 1: smartSilo1, smartSilo2, smartSilo3, smartSilo4 and smartPipe. Each one of these has a well defined interface through which it exposes its behavior to be used by the liqueur production process. The smartSilo $i$ has an input valve $INi$ and an output valve $OUTi$ through which is cyclically filled and emptied with liquid. It also has a sensor $Ei$ for the lower level and a sensor $Fi$ for the upper level. Smart silos 2 and 4 have a resistance $Ri$ to heat the liquid and a sensor $Ti$ to monitor the temperature. Smart silos 3 and 4 have a mixer $Mi$ to mix their content. Low level details as the above are encapsulated by the smartSilo to offer services of higher layer such as fill, empty, heat and mix. Silos are reserved in couples for the production of specific types of liqueur; silos 1 and 4 form one couple, silos 2 and 3 form the other couple. Raw liquid undergoes a basic process in smartSilo1 and then it is poured into smartSilo4 where it is further processed, i.e., it is heated and then mixed. Raw liquid is heated in smartSilo2 until a given temperature is reached and then it is transferred to smartSilo3 where it is mixed for a given time. The two liqueurs may be generated independently and in parallel with the constraint to use the smartPipe in an exclusive manner. Moreover, mixing the liquid in silos smartSilo3 and smartSilo4 at the same time is not permitted due to a constraint in power consumption.

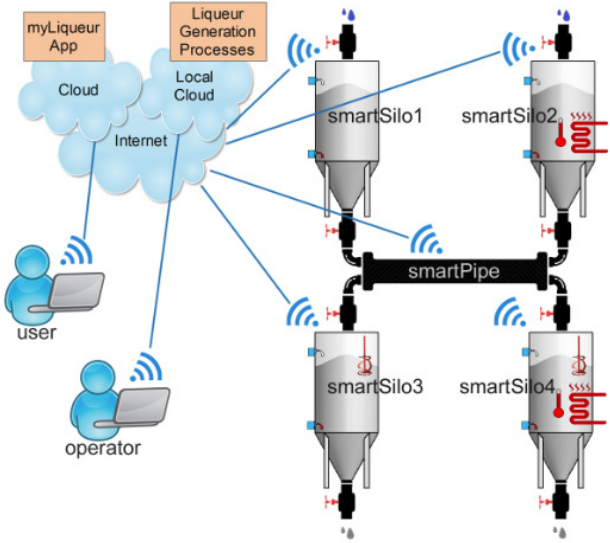

Fig. 1. The myLiqueur production system used as a case study.

### B. The motivation for the UML4IoT approach

The interface of a traditional mechatronic component such as smartSilo1 is described by a set of provided and required interfaces as shown in Fig. 2 using the UML notation. The SmartSilo class (a) implements (<<realizes>> stereotype) the SmartSiloUsageIf, which specifies the functionality provided by the mechatronic component, and (b) uses (<<use>> stereotype) the functionality specified by the SmartSiloUserIf, which specifies the functionality that a client of the component has to offer for the functionality of the SmartSilo to be effectively utilized.

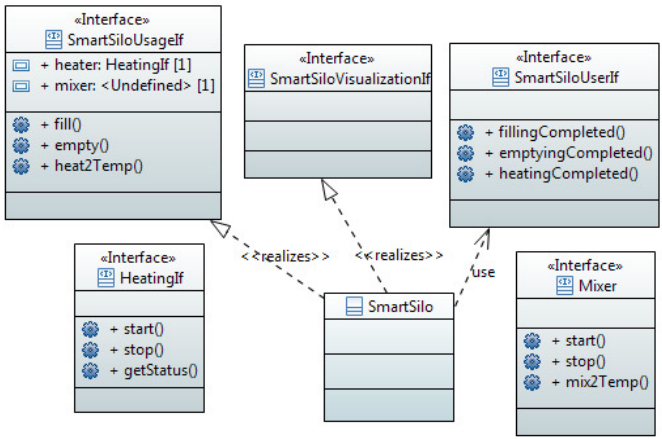

Fig. 2. Interface specification of a mechatronic component based on the OO approach.

The mechatronic component captures the low-level control (control loops) of its physical part, i.e., itsPhUnit shown in Fig.1, which imposes stringent real-time constraints not addressed by current IoT technologies. IoT technologies have to be further investigated, as also claimed in [17], regarding the requirements for reliability and real-timeliness imposed by this level of control. This is why the mechatronic component is considered as Thing in the UML4IoT approach in a similar way to the physical Mashups described in [22]. However, for the mechatronic component to be considered as Thing, a software layer is required to transform its traditional object-oriented interface, expressed with UML provided and required interfaces, to a REST based IoT-compliant interface. This layer is referred in this work as the IoTwrapper. Thus, an IoT wrapper should be added for a traditional/legacy mechatronic component to be transformed to an IoT-compliant, able to be integrated in the modern IoT environment.

IPV6 was adopted since web technologies are adopted as glue not only at the higher layers of the automation pyramid but even among the mechatronic components that constitute the manufacturing system. It is believed that IPV6-based IoT will change manufacturing leading to faster time to market, improved asset utilization and optimization. Factories and plants that are connected to the Internet will be more efficient, productive and smarter than their non-connected counterparts [23]. This is why the EU has funded several projects in this direction. Authors in [24], reporting in the context of such a project, argue that the industrial interest in manufacturing for IoT arises from its promise "to simplify initialization and reconfiguration tasks, reduce the complexity of the tasks performed by humans and lead to faster response times for the adaptations required, while at the same time minimizing configuration errors and the associated system downtime."

The LWM2M protocol [18] and the IPSO smart object [19] have been adopted for the development of the IoT wrapper to address the interoperability requirements at this level of integration. In the first prototype implementation of the case study the leshan implementation [25] of the LWM2M was used to develop the IoT wrapper. This wrapper transforms the component into an Industrial Automation Thing. Leshan is part of the IoT project of Eclipse; it is a Java implementation of the OMA LWM2M which relies on the Eclipse IoT Californium project for the CoAP and DTLS implementation. However, developing the IoT wrapper using leshan was not an easy task. A good understanding of the REST architectural paradigm and the LWM2M protocol is required along with expertise in Java programming; all these are not common skills of industrial automation engineers. This was the motivation for the UML4IoT approach presented in this paper. The approach automates the generation process of the IoT wrapper and describes the infrastructure that is required for the construction of the wrapper. The user is not required to know about REST and LWM2M, not even Java programming. He/she only has to use a UML profile, the UML4IoT, to annotate the interface of the mechatronic component and this is all that is required for the generation of the cyber part of the IoT-compliant mechatronic component. An alternative is also described for the case the UML design specification is not available. In this case the annotations can be defined at the source code specification of the cyber part of the mechatronic component. Optionally and based on requirements, IoT can be used to integrate on the mechatronic component, IoT compliant parts such as sensors and actuators in the case that real-time constraints at this level are not hard.

To the best of our knowledge there is no other work that: a) presents an approach to automate the integration process of legacy cyber-physical components in modern IoT environments, (b) examines the influences of the introduction of the IoT into the development process of the manufacturing systems, and b) presents an approach to automate the construction of a REST based IoT interface for a complex industrial automation component to transform it to an IoT-compliant smart object.

*C. The architecture of the mechatronic component*

Fig. 3 presents in SysML notation the architecture of the mechatronic component using as example the SmartSilo of the case study. The mechatronic component is composed of its physical part, i.e., *itsPhUnit* and its cyber part, i.e., *itsCyberPart*. The cyber part is further decomposed into: a) the software part (*itsS-part*), which represents the software required to transform the physical unit, i.e., the physical silo, into a smart unit, i.e., the smartSilo, and b) the electronic part (*itsE-part*), which represents the computational node required to execute the software part. The software part which is next referred as the cyber part of the smart object is further decomposed depending on its complexity to a number of classes among which we discriminate: (a) the SiloDriver, that is the software representative of the physical unit into the software domain (*itsSR*), (b) the SiloController which captures the low level control of the physical unit (*itsController*), (c) an entity object to capture the static properties of the physical unit as well as the ones of the smart object, etc. In any case all this structure is encapsulated in the mechatronic component.

SysML ports are used to represent the interaction points of the mechatronic component with its environment. A detailed description of the adopted in this work architecture is presented in [20]. Interfaces of the constituent components are an essential part of the architecture specification of the system. For the specification of the provided interface of the mechatronic component using the object oriented approach, three approaches, that have to be handled in a different way during the automatic generation of the IoT wrapper, are identified: (a) the method approach, (b) the reference approach, and (c) the hybrid approach.

*The reference approach:* The functionality of a constituent component of the smart object is exposed through a reference which is of the type of the corresponding component. In the case of SmartSilo, as an example, the complete heating functionality could be exposed by exposing the reference itsHeater of type HeaterIf. The interface HeaterIf is public so any client of the smart silo may access the heating functionality through it. In Fig. 2 this is shown by adding the attribute heater of type HeatingIf in the SmartSiloUsageIf.

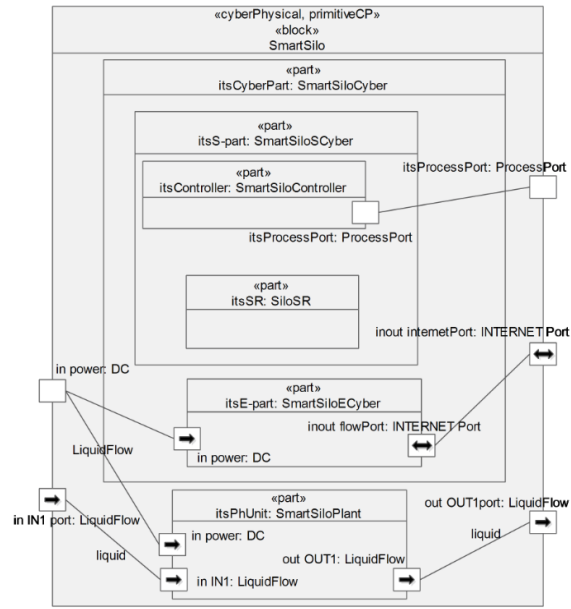

Fig. 3. The architecture of the mechatronic component.

*The method approach:* The constituent components (parts) of the smart object are encapsulated in the top level class, i.e., SmartSilo, and the whole functionality of the mechatronic component is exposed through methods of the top level class. For example, in the case of SmartSilo the complete heating functionality of its heater constituent component is exposed through the following methods of the SmartSilo class: heaterOn(), heaterOff(), getHeaterStatus(). The type Heater is private, thus no client of the smartSilo may access the heating functionality through it. Extra functionality related to heating, such as heat2Temp() may be also exposed as an extra method of the whole.

*The hybrid approach:* The functionality of the mechatronic component is exposed using both, methods and references. Consistency should be guaranteed by the whole. In the case of smart silo, as an example, the complete heating functionality is exposed by exposing the reference itsHeater of type HeaterIf and a higher layer of heating functionality is exposed through methods, e.g. heat2Temp() as shown in Fig. 2.

### III. UML4IoT - A UML PROFILE FOR IOT

The cyber part of the mechatronic component has not only to offer the services of the component to the environment but it also has to support the management of the mechatronic component, its monitoring and configuration, as well as its maintenance and repair. Interoperability is also a key requirement. OMA has developed the LWM2M standard [26] to address general requirements as the above that exist in various domains such as smart energy, manufacturing, automotive, building automation etc. The LWM2M is an application layer communication protocol that offers a standardized interface to decouple system components adopting a plug-and-play approach [18]. It is defined on top of the Constrained Application Protocol (CoAP) with UDP and

SMS bindings; the datagram transport layer security (DTLS) can be used when security for transport layer is required [26]. CoAP was developed for the M2M market with the objective to create an alternative to HTTP for RESTful APIs on resource-constrained devices [27].

The LWM2M was adopted for the mechatronic component to benefit from this communication infrastructure. LWM2M defines a server and a client to support M2M interactions. We have embedded the client part of the LWM2M in the mechatronic component to support, except from its specific functionality, general component functionalities such as discovery and registration, as well as component and service management. In order to utilize the exposed by the component functionality, other components, such as the liqueur generation processes, have to implement the server part of the protocol. It should be noted that even though the mechatronic component is equipped with the client part of the protocol, it is the actual provider of the component's services to the environment. The interface of the LWM2M is defined on top of an extensible object model; it is based on the REST architectural paradigm and satisfies requirements regarding performance and constraints of M2M devices. The resource is the key concept of the REST paradigm., Any static or dynamic property of the mechatronic component that has to be exposed should be considered as a resource. Fields and operations of the smart silo and the references of its provided interfaces should be handled as resources. Resources are organized into objects, with an object type to define the logical organization of resources as shown in the UML diagram of Fig. 4, which captures the core constructs of the LWM2M protocol used for the definition of the UML4IoT profile.

LWM2M defines four interfaces: (a) bootstrap, (b) client registration, (c) device management and service enablement, and (d) information reporting. The device management and service enablement interface supports access to object instances and resources on the mechatronic component, while the information reporting interface supports asynchronous notification based on corresponding subscriptions. Fig. 4 presents also the operations that are supported for the core constructs of the LWM2M object model. The Execute operation is used to initiate some action and can only be used on a Resource. The Create and Delete operations of the device management and service enablement interface are used to create and delete object instances. All other operations, i.e., Read, Discover, Write and WriteAttributes, may apply on Resource, Resource Instance, Object and Object Instance.

The object model of the LWM2M can be used to define the structure of the information that is exposed by the mechatronic component as well as the operations that may be applied on this information. For a very simple mechatronic component this model is appropriate to express also the structure of its cyber part. However, if the cyber part implements control and coordination logic that is usually required by its physical part then the LWM2M object model is not appropriate to define its structure. In this case the model of the cyber part is constructed following the traditional OO approach and UML is used to represent it. Except from complexity, there is another reason for using this approach. Legacy systems are already specified in UML or at the level of the source code and their exposed functionality is defined in terms of provided and required interfaces. This is why a mapping is proposed in this work of the UML traditional OO interface specification to the LWM2M-compliant REST interface. This mapping allows for the automation of the transformation process of the UML traditional OO interface to a REST-like interface and more specifically to a LWM2M-compliant one. The basic idea for the automation of this transformation process is the use of a UML profile.

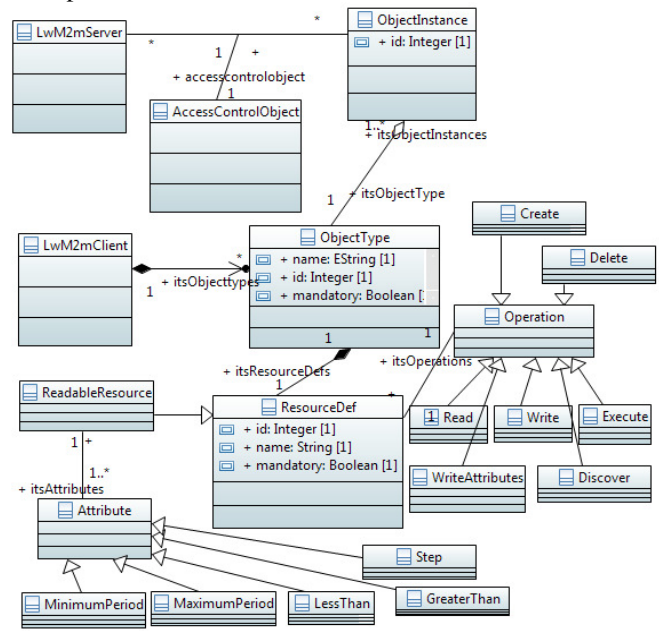

Fig. 4. Part of the OMA LWM2M object model that is the base for the definition of the core constructs of the UML4IoT profile

The profile is the lightweight extension mechanism provided by UML to allow the extension and specialization of its meta model with constructs that are specific to a particular domain. UML profiles have been already used in the domain of embedded and real-time systems, as for example [28][29]. Authors in [30] review the most important UML profiles for real-time systems and the research activities around these profiles. In this work, UML meta-classes, such as Class, Property and Operation, are extended and specialized to represent basic constructs of the REST paradigm to facilitate the transformation process of the UML traditional OO interface to a REST-like one. The UML4IoT profile is used to annotate, on the UML model of the cyber part of the mechatronic component, those artifacts of the model that represent exported properties of the component. The ObjectType stereotype extends the Class artifact as shown in Fig. 5, which presents the core part of the UML4IoT profile. It defines the LWM2M object as a composition of LWM2M resources modeled by the Resource stereotype.

The Resource stereotype is the generalization of three other stereotypes, two of which extend the Operation metaclass, i.e., the Operation resource and the InstanceResource, and one, the

ObservableResource, which extends the Property metaclass. The ObservableResource stereotype has been defined to annotate any property of the mechatronic component for which there is a need of utilizing the notification interface of the LWM2M protocol. The ObjectType stereotype is used to annotate the class that represents the cyber part of the mechatronic component as well as any other class or interface that classifies the attributes exposed by the provided interface of the cyber part of the component.

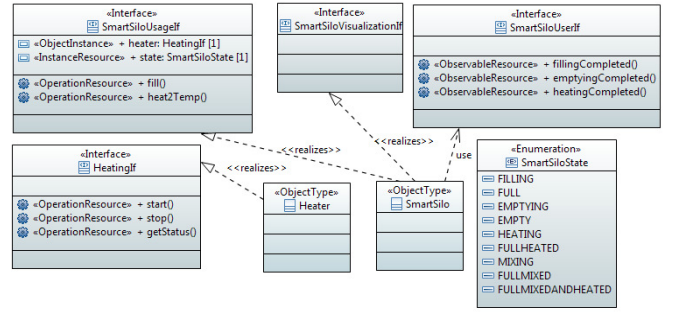

Fig. 6. The cyber interface of the SmartSilo annotated with the UML4IoT profile.

## IV. AUTOMATING THE GENERATION OF THE IoT WRAPPER

An alternative that can be adopted when the UML design specification is not available is to directly annotate the cyber part of the component on the source code. In [31] the application of the UML4IoT in the case that the cyber part is developed using the IEC 61131 function block model, which is widely used in industry, is described. This allows the wrapping of legacy IEC 61131 based components with an IoT REST-like interface that allows these to be integrated in the modern IoT manufacturing environment.

In this work the application of the UML4IoT profile using Java as implementation language for the cyber part of mechatronic component is described. Java was selected since it supports through the mechanism of reflection meta programming that allows a fully automated generation of the IoT wrapper from the annotated source code. The Java annotations required for the annotation of the source code are firstly described and then their application and exploitation towards the generation of the IoT wrapper.

### A. The Java LWM2M Annotations

The Java LWM2M annotations have been defined using as base the UML4IoT profile. Fig. 7 presents the ObjectType and the Resource annotation definitions which are part of the lwm2m package of the UML4IoT java implementation. The `ObjectType` and `Resource` annotations are used to annotate the object types and the resource types of the java code that would be utilized for the construction of the SmartSilo json file, that contains the descriptions of objects and resources required by leshan. The ObjectInstance annotation is used for the partial generation of instanceEnablers that are key part of the leshan-based specification of the IoT wrapper of the mechatronic component.

Fig. 8 present part of the cyber part of the SmartSilo source code enriched using the lwm2m package annotations in the form of annotation instances. Only the exposed properties of the SmartSilo are shown in this figure. Not exposed properties or methods are not annotated. As shown, object types and resources have been annotated using the set of REST interfaces defined by IPSO [32], which results to IPSO-compliant IoT wrappers. IPSO to enable interoperability between heterogeneous components has also defined in the IPSO Smart object Guideline [19][33] a set of standard object

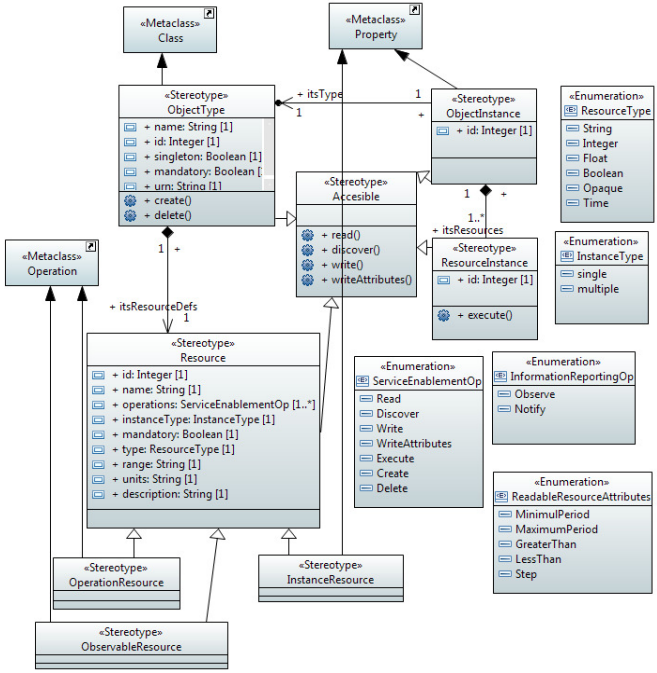

Fig. 5. The core part of the UML4IoT profile for LwM2M

The UML4IoT profile, which was developed using Papyrous, is used to annotate the interface of the cyber part of the mechatronic component. As an example, Fig. 6 presents the part of the class diagram of the cyber part of the SmartSilo that captures the cyber interface of the SmartSilo, which has to be exposed, annotated with the UML4IoT profile. The references of the provided interface are annotated with the <<ObjectInstance>> stereotype, while their types are annotated with the <<ObjectType>> stereotype. All the methods of the provided interfaces, i.e., SmartSiloUsageIf and HeatingIf are annotated with the <<OperationResource>> stereotype. Methods of the required interfaces such as the SmartSiloUserIf are annotated with the <<ObservableResource>> stereotype.

The UML4IoT profile may be used when the UML design specification of the cyber part of the mechatronic component is available or when this can be generated using reverse engineering from the source code. In this case the designer may properly annotate the exposed properties of the cyber part of the component using a UML tool. Using the code generation functionality of the UML tool the IoT annotations are transformed to the source code.

7types along with their exposed resources. Types defined by IPSO include among others Temperature Sensor, Actuation, Presence Sensor, Light Control, etc. For each smart object, IPSO defines the objectID and the resources that it exposes. For example for the smart object Temperature sensor has id 3303 and Sensor Value, units, minMeasuredValue, maxMeasuredValue, minRangeValue, maxRangeValue and, Reset Min and Max Measured Values, as resources. Each resource has predefined properties such as id, type and access type. As shown in figure 8 custom ids have been defined for the type and resources that are not supported by the IPSO.

```java
public @interface ObjectTypeAno {
        public String name();
        public int id();
        public InstanceTypeEnum instanceType();
        public boolean mandatory();
        public String description();
}

public @interface ResourceAno {
        public int id();
        public String name();
        public ServiceEnablementOpEnum[] operations();
        public InstanceTypeEnum instanceType();
        public boolean mandatory();
        public ResourceTypeEnum type();
        public String range();
        public String units();
        public String description();
        }
```

Fig. 7. Example definitions of lwm2m Java Annotations

### B. Implementation alternatives

The enriched with the lwm2m annotations java code can be exploited for the generation of the IoT wrapper in three different approaches: (a) the edit-time approach, (b) the load-time approach, and (c) the run-time approach.

Based on the edit-time approach the annotated source code is used to automatically generate during edit time the infrastructure, i.e., the json file and skeleton code of the source code, which are required for the generation of the IoT wrapper using leshan.

The other two approaches are based on the transformation of the annotations at the compile time from the source code to the java bytecodes. In this way this information is available at load and run-time. It is estimated that the run-time approach will introduce a high performance overhead on the mechatronic component, thus, it was decided to proceed with the load-time one.

A prototype implementation of the load time approach which focus on service enablement was developed and is used for demonstration and performance evaluation. Based on this approach annotations are used at class load time and are exploited through the use of the java reflection mechanism to implement LWM2M. This approach was adopted as more powerful and promising for a completely flexible and automated process for the generation of IoT wrappers for smart objects. The approach introduces an extra overhead compared to the leshan implementation, but it leads to a more flexible and effective implementation of the LwM2m protocol.

## V. RELATED WORK

CPSs play an important role towards Industry 4.0. The great impact of CPSs in manufacturing based on a number of explorative case studies is examined in [6]. Authors argue that CPSs are transforming the service business in manufacturing and offer new opportunities for business innovation. Real-time requirements on manufacturing systems as this regards the adoption of the CPS concept in their development are discussed in [34]. Authors propose the use of Ethernet and CAN-based real-time communication protocols and describe a three layered software architecture which they propose for addressing self-reconfiguration. In the UML4IoT the low level control of the physical unit is encapsulated into the MTC to allow the vendor to use its proprietary technology for its implementation. This implementation will be hidden from the environment since the mechatronic component appears with an IoT-compliant interface.

The current status of cyber-physical systems in manufacturing is presented in [2]. Specific examples of CPS in manufacturing are presented and discussed and authors argue

```java
@ObjectTypeAno (name = "SmartSilo",id = 16663,instanceType = InstanceTypeEnum.SINGLE,mandatory = true,description = "")
public class SmartSilo {
        @ResourceDefAno (id = 0, name = "filling",operations = {ServiceEnablementOpEnum.READ},
                    instanceType = InstanceTypeEnum.SINGLE,
                    mandatory= true, type = ResourceTypeEnum.BOOLEAN, range = "", units="",description="")
        public Boolean filling;

        @ObjectInstanceAno(id=0, name = "heater", objectTypeId = 16668)
        public MyHeater heater = new MyHeater();

        @ObjectInstanceAno(id=1, name = "inValve",objectTypeId = 16664)
        public Valve inValve  = new Valve();
        . . .
        @ResourceDefAno (id = 2, name = "fill",operations = {ServiceEnablementOpEnum.EXECUTE},
                    instanceType = InstanceTypeEnum.SINGLE,mandatory= true, type = ResourceTypeEnum.BOOLEAN,
                    range = "", units="",description="")
        public void fill(){  . . . }
}
```

Fig. 8. Sample java code of SmartSilo annotated with the lwm2m annotations

that CPSs is a promising approach for factories. Among the questions authors discuss is the following: "How does the term CPS relate to other concepts such as IoT, big data and systems of systems?" In their discussion authors refer to two visions of the IoT. In the first one IoT is considered as enabling technology that can be used to develop a special class of CPS, i.e., systems including the Internet; the second vision extends IoT beyond basic communication with the ability "to link "cloud" representations of the real things with additional information such as location, status, and business related data." In UML4IoT a third vision is added. IoT technologies are used as the glue that integrates the components of the cyber-physical system, that maybe cyber, cyber-physical and human, as far as it regards their cyber interfaces. Thus, CPSs developed using as glue IoT technologies will be an integral part of the IoT since its constituent components are Things of the IoT. Of course CPSs that do not use IoT technologies for their integration may also be part of the IoT by using the IoT wrapper. In this case the CPS is the Thing. UML4IoT can be utilized in both cases increasing the productivity and the effectiveness of the development process.

UML and SysML are widely accepted as the de-facto standards for software and systems development respectively. They increase the level of abstraction in system specification and can be used as a first step towards the adoption of the model driven engineering paradigm [35]. As claimed in [36] "UML is still the first choice of practitioners for specifying software architectures," with most Architecture Description Languages mainly used in the research community. A specific use of SysML and UML for the modeling of the mechatronic component is described in [20]. UML4IoT extends this work to address also the integration at the system level using as glue the IoT.

Web protocols, such as HTTP and SOAP, have been developed for the integration of information systems and the exploitation of their services from humans. These protocols have been investigated for a long for the integration of manufacturing systems and it was found that are not appropriate for the integration of the new generation of manufacturing systems where machine to machine communication is a key issue. Authors in [37], ten years ago, described opportunities and challenges in using the service oriented architecture in manufacturing. Since then several research articles published reporting successful or promising results regarding the exploitation of the SOA paradigm in the industrial automation domain, e.g., [38-40]. SOAP has been defined as a lightweight protocol intended for exchanging structured information in a decentralized, distributed environment [41]. However, SOAP is today not the preferred technology for the IoT; the REST architectural paradigm [42] is considered as the dominating one [20]. The appearance in the market, during past years, of various PLCs with embedded HTTP servers was the motivation for the analysis of the overhead introduced by the HTTP in manufacturing. Authors in [43] found that the use of HTTP at the device level is introducing performance overhead that allows the approach to be considered only for soft real-time systems. The performance of PLC-to-PLC communications based on HTTP is evaluated in [44] and it is compared to Modbus TCP. Authors argue that these PLCs may be used in collaboration with PLCs that acts as the HTTP clients, to allow the integration of control systems with soft real-time constraints. Authors also claim that while SOA's suitability is proven in IT systems, it has not been adopted yet in commercial PLCs, and thus cannot be considered as a solution for integration with already deployed control systems. They attribute this result mainly to the relatively low performance of PLC application code executing complex string processing required by the HTTP protocol. The HTTP communications is considered as an alternative that is worth evaluating for soft real-time NCS.

IoT has already attracted the interest of the research community in automations systems and manufacturing. Authors in [3] investigate the impact of IoT in modern manufacturing and argue that the emerging IoT infrastructure is able to support effectively the information systems of the next-generation manufacturing enterprises. In UML4IoT the IoT is effectively used to support not only the information systems of manufacturing but it plays a leading role in the integration of all constituent components of a modern manufacturing system, which are cyber, cyber-physical and human. Authors in [24] describe, as result of an FP7 EU project, the impact of IoT on factory automation and claim that factory automation could benefit from IoT by making the manufacturing environment more agile and flexible. Authors refer to eight high-importance general requirements for manufacturing systems and very abstractly describe an IoT-centered architecture with main objective to allow an IoT compliant management of devices and services, which satisfy requirements and constraints of manufacturing environments such as the requirements for reliable communication and guaranteed security. They do not refer to any specific IoT technology and do not describe a concrete way of using IoT at the production infrastructure layer. Moreover, they allocate controller logic at the Cloud computing environment layer.

It is widely accepted today that manufacturing is slowly but steadily experiencing a paradigm shift [45][46] towards what is known as Industry 4.0. This is why big players in the IT such as AT&T, Cisco, General Electric, IBM, and Intel initiated a not-for-profit, open membership organization, the Industrial Internet Consortium (http://www.iiconsortium.org/) to coordinate the priorities and enabling technologies of the Industrial Internet. The objective is to improve properties of CPS such as openness, autonomy, distributed control, adaptability, discipline integration, etc. An extensive list of the properties of manufacturing systems that can be improved adopting current trends in IT is given in [2]. Cloud manufacturing has also emerged as a new manufacturing paradigm where timely process planning can be assisted by real-time monitoring of both the availability and status of machines and this unlocks business opportunities toward service-oriented manufacturing [47].

## VI. MEASUREMENTS AND EVALUATION

To evaluate the timing behavior of the IPSO-compliant mechatronic component and the overhead introduced by the IoTwrapper, a number of measurements have been performed using as test bed the prototype implementation of the liqueur production system. Two implementations of the IoT wrapper are used in the measurements.

In the first deployment scenario the wrapper has been developed using the leshan implementation of the OMA LWM2M. In this case the IoTwrapper, i.e., the leshan wrapper, was generated manually using the traditional method proposed by leshan. The Json file was generated automatically from the annotations of the java source code. Annotations of the source code were also used to speed up the development process of the IoT wrapper. This process is estimated that can be semi-automated and is a work in progress. In the second scenario the IoT wrapper, i.e., the UML4IoT wrapper, is developed in a fully automated manner by just compiling the annotated java source code of the cyber part of the mechatronic component and importing in the project the UML4IoT implementation of the LWM2M.

For both implementations three run-time configurations, all based on a 100Mbps LAN, have been used to measure the round-trip time for each one of the EXECUTE and READ operations of LWM2M. The three run-time configurations differ on the computation node on which the liqueur generation process is deployed. In the $1^{st}$ configuration, the liqueur generation process is deployed on the computation node of the smartSilo, i.e. Raspberry Pi; in the $2^{nd}$ on the PC of the local LAN; in the $3^{rd}$ on the public Cloud. Measurements do not include the operation execution time; it is a measurement between the time the operation is issued from the LWM2M server to the time the response of the LWM2M client to this command is received back to the server.

The characteristics of the three computational nodes used in the liqueur production prototype system for measurements are the following:

a) Raspberry Pi : The mechatronic component is equipped with a Raspberry pi model B+ board with a 700-MHz 32bit ARM1176JZFS CPU, 512-MB of RAM and a microSD memory card running linux debian 7 with java hotspot client 25.0-b70 JVM installed.

b) PC: The PC is used for the execution of the liqueur generation process. It is equipped with an AMD athlon II X2 235e CPU running at 2.7GHz and 4 GB of DDR3 RAM, Windows 7 64bit OS, java hotspot client build 25.65-b01 JVM installed, and

c) A virtual PC: This computation node that is used as an alternative for the execution of the liqueur generation process in public cloud, was created on Okeanos, a cloud service for the Greek Research and Academic Community (https://okeanos.grnet.gr/home/). It has two QEMU virtual CPUs version 2.1.2 at 2.1GHz and 6GB of RAM running windows server 2012 and java hotspot 64-bit server build 25.40-b25.

For each one of the three run-time configuration, 1,000 EXECUTE or READ operations were executed for each one of the two wrappers, i.e., the leshan and the UML4IoT one. Table I presents in milliseconds the min, max, average and standard deviation for every scenario for the two wrappers regarding the EXECUTE operation. The leshan wrapper is faster compared to the UML4IoT but this was expected since the use of metaprogramming introduces performance overhead in the LWM2M implementation. This is the cost that we have to pay for getting the high flexibility and the full automation of the generation process of the IoT wrapper. From the measurements it is also clear that the LWM2M IoT protocol stack and the specific implementation, i.e., leshan, is not appropriate for real time operations since it introduces an average of 3.02 millisecond for a round trip in an EXECUTE operation with a possible high up to 62.36 ms, a time that is not accepted in manufacturing control systems. This proves our decision to capture low level control of the physical unit inside the corresponding mechatronic component and allow the developer of the component to use its own communication protocol if one is required for the integration of its constituent parts or components in the case of a composite component. One may also note that the average round trip measured for the $1^{st}$ scenario is higher compared to the $2^{nd}$ one that includes the local LAN. This is reasonable since the PC is faster compared to the Raspberry Pi as shown also from the average round-trip time over 1,000 READ operations for the leshan wrapper that is 1.89 milliseconds with min 1.44 and max 12.87 ms, when both the smartSilo cyber part and the liqueur generation process are deployed on the PC.

Table II is for the READ operation. Fig. 9 presents the distributions of the measurements for the READ operation on the $2^{nd}$ scenario.

TABLE I
TIMING CHARACTERISTICS (IN MS) FOR THE EXECUTE OPERATION

| EXECUTE | leshan | | | UML4IoT | | |
|---|---|---|---|---|---|---|
| | 1st Conf. | 2nd Conf. | 3rd Conf. | 1st Conf. | 2nd Conf. | 3rd Conf. |
| min | 2.61 | 2.04 | 22.19 | 4.64 | 2.37 | 22.23 |
| max | 6.12 | 3.64 | 44.95 | 8.69 | 5.43 | 39.22 |
| avg | 2.78 | 2.30 | 24.25 | 4.95 | 2.62 | 24.01 |
| stddev | 0.17 | 0.16 | 1.75 | 0.34 | 0.16 | 1.16 |

TABLE II
TIMING CHARACTERISTICS (IN MS) FOR THE READ OPERATION

| READ | leshan | | | UML4IoT | | |
|---|---|---|---|---|---|---|
| | 1st Conf. | 2nd Conf. | 3rd Conf. | 1st Conf. | 2nd Conf. | 3rd Conf. |
| min | 2.75 | 1.92 | 22.61 | 4.75 | 2.44 | 22.84 |
| max | 6.43 | 4.09 | 46.91 | 8.60 | 4.12 | 41.03 |
| avg | 2.92 | 2.35 | 24.44 | 5.04 | 2.71 | 24.66 |
| stddev | 0.25 | 0.19 | 1.47 | 0.33 | 0.15 | 1.39 |

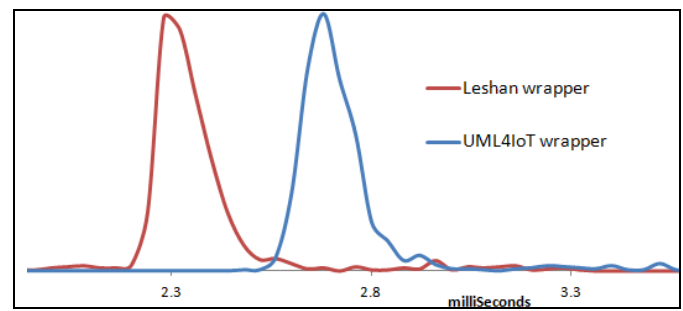

Fig. 9. Distributions of 1,000 READ operations for the $2^{nd}$ run-time configuration for the leshan and UML4IoT based wrappers.

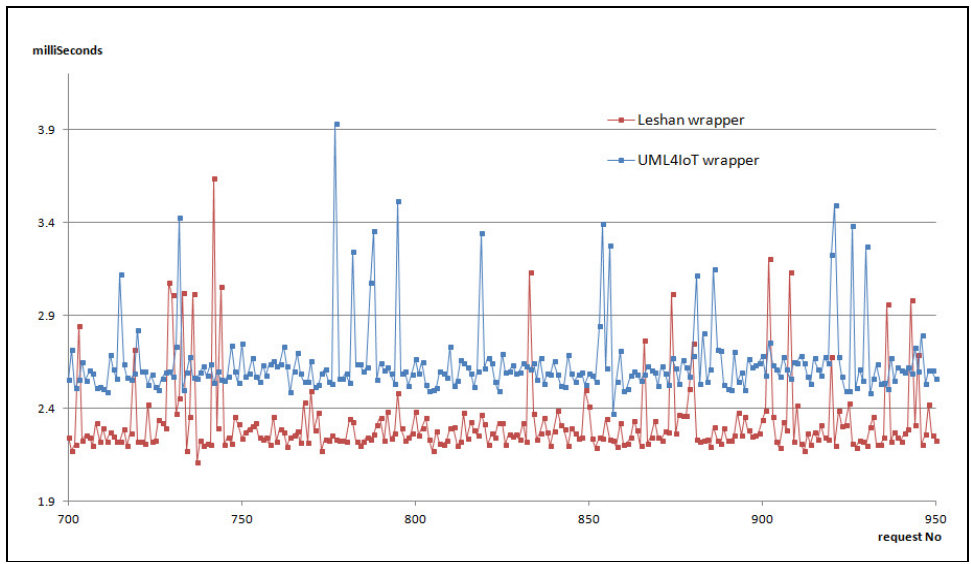

Fig. 10. A sample of measurements from the set of 1,000 EXECUTE operations for the 2nd run-time configuration for the leshan and UML4IoT based wrappers.

The extra overhead introduced by the UML2IoT wrapper is evident. Fig. 10 presents a sample of measurements from the set of 1,000 EXECUTE operations for the 2nd scenario. It is also interesting to note that for the case that Cloud is included in the path then the performance overhead introduced by the UML4IoT wrapper is negligible compared with the leshan one.

## VII. CONCLUSION

IoT is transforming the way that modern manufacturing systems will be developed and operate. As expected the introduction of this new technology influences the development process by introducing the REST architectural paradigm. It imposes a paradigm shift for the automation system developer and requires effective approaches to handle the complexity in this transition. Moreover, there is a need for legacy manufacturing components to be integrated in the modern IoT manufacturing environment. In this paper an approach is described to address these challenges. A UML profile for IoT (UML4IoT) is defined to allow the developer to automatically generate the IoT-compliant interface of the mechatronic components and the implementation of the corresponding wrapper. An alternative is also defined for the case that a UML design specification is not available. The properties of the mechatronic component that should be exposed are annotated on the source code of its cyber part and the resulting code is used to automatically generate the layer that should wrap the component to present an IoT-compliant interface. Both approaches may be used in the generation process of new components but also in bringing legacy components in the modern IoT manufacturing environment.

The prototype implementation of the myLiquer laboratory system has proved the effectiveness of the UML4IoT approach and demonstrates its applicability. Even though a partial implementation of the LWM2M that supports only the service enablement interface has been developed at the time, the comparison with the leshan implementation regarding performance is an indication that the approach is very promising since it supports a fully automated generation of the IoT wrapper with a small cost in performance. Our plans include (a) the implementation of other key interfaces of the LWM2M, (b) the implementation of a transformer to utilize the edit time annotations to semi automate the generation of the IoT wrapper based on leshan and (c) improve the application of the approach for the case that the IEC 61131 function block is used for the specification of the cyber part of the mechatronic component. The integration of UML4IoT with the leshan implementation is estimated that would offer an optimal solution in terms and performance and flexibility.

ACKNOWLEDGMENTS

Authors would like to thank the leshan development team and more specifically Simon Bernard and J.F. Schloman for the support on using leshan.